# Atomically-thin Ohmic Edge Contacts Between Two-dimensional Materials

*Marcos H. D. Guimarães[‡ 1,2], Hui Gao[‡ 3], Yimo Han[4], Kibum Kang[3], Saien Xie[4], Cheol-Joo Kim[3], David A. Muller[1,4], Daniel C. Ralph[1,2], and Jiwoong Park*[1,3]*

[1] Kavli Institute at Cornell for Nanoscale Science, Cornell University, Ithaca, New York 14853, USA

[2] Laboratory of Atomic and Solid State Physics, Cornell University, Ithaca, NY 14853, USA

[3] Department of Chemistry and Chemical Biology, Cornell University, Ithaca, New York 14853, USA

[4] School of Applied and Engineering Physics, Cornell University, Ithaca, NY 14853, USA

**ABSTRACT**

With the decrease of the dimensions of electronic devices, the role played by electrical contacts is ever increasing, eventually coming to dominate the overall device volume and total resistance. This is especially problematic for monolayers of semiconducting transition metal dichalcogenides (TMDs), which are promising candidates for atomically thin electronics. Ideal electrical contacts to them would require the use of similarly thin electrode materials while maintaining low contact resistances. Here we report a scalable method to fabricate ohmic graphene edge contacts to two representative monolayer TMDs - $MoS_2$ and $WS_2$. The graphene and TMD layer are laterally connected with wafer-scale homogeneity, no observable overlap or gap, and a low average contact resistance of 30 k$\Omega\mu$m. The resulting graphene edge contacts show linear current-voltage (IV) characteristics at room temperature, with ohmic behavior maintained down to liquid helium temperatures.

**Introduction**

Even though monolayer two-dimensional (2D) materials[1,2] such as graphene[3] and transition metal dichalcogenides (TMDs)[4] show promising results towards atomically-thin circuitry[4–11], the contact volume and resistance often dominate over the total device volume and resistance. There are two fundamentally different contact interface geometries for 2D materials[12,13]: top contacts and edge contacts (Figure 1a). Conventional methods use 3D metallic electrodes to top-contact monolayer 2D materials. Recent developments have shown that low contact resistance is achievable in this configuration[13–23], but the total electrode volume is an intrinsic problem for this approach. Graphene top-contacts can provide much smaller volumes with low contact resistances when they have sufficiently large contact areas[24,25]. However, due to the van der Waals gap[13,26,27] between graphene and the TMD, the contact resistance increases dramatically as the length of the graphene top contact is reduced below the transfer length to the tens of nm scale[28].

Edge contacts, on the other hand, offer the potential for efficient carrier injection to atomically thin materials despite a much smaller contact area defined by their atomic thickness[26]. Conventional metal electrodes have been successfully used to make edge contacts to graphene[29], but the large electrodes still dominate the device volume. Another approach, which alters the crystalline phase of a 2D TMD semiconductor to make it metallic, generates an edge contact to the TMD with small contact volume and resistance[30–32]. However, it relies on a phase that is metastable, and it uses methods that are customized for the specific chemical composition of the TMD.

To realize the full potential of atomically-thin TMD materials for electronics will require contacts with a low intrinsic volume that are scalable with low contact resistances, chemically and thermally stable, and versatile towards use with different TMD materials. Here we demonstrate edge contacts to monolayer TMDs that fulfill these requirements by using monolayer graphene as the electrode. We fabricate laterally-stitched graphene/TMD heterostructures using a scalable and patternable growth method with homogeneous quality over the entire substrate. The resulting one-dimensional graphene (1DG) contacts show low contact resistance ($R_c$) of approximately 30 kΩμm, with ohmic behavior down to liquid helium temperatures, while adding the least possible additional volume to the devices.

**Results and Discussion**

Figure 1b summarizes our approach. A monolayer graphene film, grown by chemical vapor deposition (CVD)[33] is transferred onto a $SiO_2$/Si substrate and patterned by photolithography and oxygen plasma etching. The key step for making lateral connections between the graphene and a TMD is the use of a highly controllable Metal-Organic CVD (MOCVD) method[34] for the growth of single layer TMDs ($MoS_2$ or $WS_2$) from the graphene edges, causing the ensuing TMD growth to occur only on the exposed $SiO_2$ surface and not on the graphene (see Methods and Supplementary Information for further details). The resulting heterostructure film is then further processed to fabricate an array of devices that use graphene as one-dimensional edge contacts to the TMD channels.

In Figure 1c we plot the gate-dependent 2-probe conductance of a representative $MoS_2$ device (22 μm long) contacted by graphene electrodes (20 μm wide) and we compare it to the results from a device with conventional metal top (2DM) contacts with dimensions 23 μm ✕ 22 μm ✕ 55 nm. All results shown here and discussed below were obtained at room temperature and ambient conditions, unless otherwise noted. Both sets of contacts show ohmic behavior (see inset).

Surprisingly, the graphene electrodes show a lower contact resistance despite drastic reduction in the electrode volume and, as we confirm below, the contact area. This results in a higher 2-probe conductance, and therefore also an enhanced field-effect mobility for the devices with 1DG contacts (Figure 1c; main panel). The improved contact resistance of the graphene electrodes implies that there is a strong connection between the graphene and TMD edges, suggesting a lack of a van der Waals gap or tunnel barrier as previously observed in 2DM contacts[13,16] or overlapped graphene top contacts[24,28].

The growth and fabrication process described above results in lateral TMD/graphene heterostructures with uniform properties over large areas. An optical micrograph of the heterostructure film over centimeter scales is shown in Figure 2a. We observe homogeneous graphene/$MoS_2$ heterostructures over the entire substrate with no visible overlap or multilayer regions in the optical contrast (see the zoomed-in image in the inset, Figure 2a). The monolayer homogeneity and spatial controllability are further confirmed by the $MoS_2$ photoluminescence (PL; Figure 2b) and graphene G-band Raman mapping (Figure 2c). The PL and Raman intensity profiles extracted along the dashed lines of Figure 2b and 2c show abrupt transitions within the optical resolution (approximately 1 μm) between the two materials (Figure 2d) demonstrating excellent compatibility with large area patterning with precise spatial control. The TMD layer show good optical properties as demonstrated by the sharp peaks in their PL spectra, similar to the ones obtained in exfoliated flakes[35,36] (Figure 2e and 2f).

Systematic studies further show that the nucleation behavior of the TMD is strongly dependent on the partial pressure ($P_M$) of the transition metal precursor ($Mo(CO)_6$ for $MoS_2$ and $W(CO)_6$ for $WS_2$) during the growth. A scanning electron microscopy (SEM) image of a representative sample grown under optimized conditions, with $P_M$ below 0.7 mTorr, is shown in Figure 3a (main panel) where we observe no $MoS_2$ nucleation on graphene nor the formation of multilayer $MoS_2$ regions in our TMD film. On the other hand, when the growth is performed under a more reactive environment (higher $P_M$), we observe multilayer $MoS_2$ regions on both the graphene and the $MoS_2$ film (Figure 3a, inset). As shown below, the nucleation behavior of the TMD has a direct impact on the lateral stitching of the graphene and the TMD.

The lateral connection between the graphene and the TMD can be probed by dark-field electron microscopy (DF-TEM). Figure 3b shows a representative DF-TEM image of a graphene/$MoS_2$ junction grown under optimized conditions, showing no overlap region between the graphene and $MoS_2$ within the imaging resolution (below 10 nm). The selected diffraction spots are shown on the upper right inset. The lateral connection between graphene and $MoS_2$ is observed consistently over different regions of the substrate (see Supplementary Information for more images). A representative example of an overlapped graphene/$MoS_2$ junction, grown under high $P_M$ (above 1 mTorr), is shown in the inset of Figure 3b. Its overlapped structure is similar to the ones that were recently reported[28], which displayed much higher contact resistance (approximately 300 kΩμm) than our laterally-stitched 1DG contacts (see Figure 4c for a more quantitative comparison).

The formation of the lateral connection between graphene and $MoS_2$ (or $WS_2$) only at low $P_M$, is consistent with the layer-by-layer growth mode as in our previous report on monolayer TMD growth by MOCVD[34]. There, the nucleation and growth was limited to the $SiO_2$ growth surface until a fully continuous monolayer was formed. On the other hand, multilayer regions were found to form if the precursor concentration was higher. In our current work, we similarly found that the TMD only nucleates on the SiO$_2$ surface including at the graphene edges at lower $P_M$, and the TMD grains grow on the SiO$_2$ surface until they meet and laterally connect to form a homogeneous layer. In contrast, with higher $P_M$, the TMD nucleates also on the graphene surface, leading to multilayer formation and regions of overlapped graphene/TMD junctions (see inset of Figure 3a and Supplementary Information). Our results thus strongly suggest that the precise control of all the precursor pressures, which is a key feature of our MOCVD approach, is central to the fabrication of laterally connected edge-contacts between graphene and TMDs.

We performed quantitative determinations of contact resistances using the analog of Transfer Length Measurements (TLMs)[16,19,31], based on the 2-probe resistance of TMD channels measured with varying length and fixed width (see Figure 4a for an optical image of a device). The total 2-probe resistance is $R = 2R_c + \left(\frac{\rho_{TMD}}{W}\right)L$, where $\rho_{TMD}$ is the TMD resistivity, and $W$ and $L$ are the TMD channel width and length respectively.  Figure 4b shows the measured dependence of $R$ on $L$ for two devices with 1DG contacts to MoS$_2$ (black circles) and WS$_2$ (blue triangles) channels. These measurements were taken at a top gate voltage of $V_{TG}$ = 3 V, corresponding to a carrier density of n ~ 1 $\times$ 10$^{13}$ cm$^{-2}$, estimated from the threshold voltage and the gate capacitance. The $R_c$ values are obtained by extrapolating to zero TMD length, which are similarly low for the 1DG contacts to both MoS$_2$ and WS$_2$, on the order of 20 kΩμm. This suggests that our method is versatile towards its use with different TMDs.

The one-dimensional graphene edge contacts show consistently low contact resistances with good reproducibility. In Figure 4c we plot our results for $R_c$ obtained by TLM measurements for seven MoS$_2$ (solid circles) and two WS$_2$ based devices (solid squares) with 1DG contacts, obtained from five different growth runs on different substrates. The values for $R_c$, measured at a carrier density of n ~ 1 × 10$^{13}$ cm$^{-2}$, remain similar throughout different samples, with the median $R_c$ of 30 kΩμm. For a direct comparison, we fabricated three MoS$_2$ based devices using Ti/Au metal electrodes, which are widely used to contact TMD materials, next to some of the devices with 1DG contacts. These devices with 2DM contacts show higher contact resistance values ($R_c$ ~ 95 kΩμm; denoted by solid diamonds, Figure 4c) at similar carrier densities that are consistent with previously reported results[14]. These results confirm that our graphene electrodes provide low resistance edge contacts to TMDs despite the minimal electrode volume with $R_c$ values that are smaller than those of conventional Ti/Au 2DM contacts but are larger than the smallest values recently reported from pure metal electrodes[20,21]. Here we restrict our comparison to contacts to monolayer TMDs only, since $R_c$ is known to increase with the decrease in layer number[13,23,37,38].

Figure 4c presents additional comparison with graphene top (2DG) contacts, which have achieved similarly low contact resistances in the order of $R_c$ ~ 20 kΩμm while reducing significantly the electrode volume when compared to 2DM contacts[24,25] (open circle in Figure 4c). However, when the overlap area between the graphene top contact and MoS$_2$ is reduced, the devices show an increase in contact resistance ($R_c$ ~ 300 kΩμm; open triangle in Figure 4c) and display nonlinear IV characteristics at room temperature[28], showing that 2DG contacts may not be suitable for small scale devices. Likewise, $R_c$ in devices with 2DM contacts is known to increase exponentially when the contact length is decreased below the transfer length[16,19,38] ($L_T$ ~ 15 – 600 nm), which limits the minimum device footprint. On the other hand, our 1DG contacts should not be bound by such limitations, since it uses the edge contact geometry.

Furthermore, temperature dependent electrical measurements confirm the ohmic nature of the graphene edge contacts to $MoS_2$: our $MoS_2$ devices with 1DG contacts show linear IV characteristics with little temperature dependence from room temperature down to liquid helium temperatures (Figure 4d), and we observe no temperature dependence for $R_c$, extracted from TLM, over the same temperature range (Figure 4e). Altogether, these results suggest that 1DG contacts provide a novel route for reducing the overall device size while maintaining low-resistance ohmic contacts.

For doping-dependent studies, we have performed direct measurements of $R_c$ using a gated 4-probe geometry for junctions in devices with smaller dimensions (Figure 5a, bottom inset). For the 4-probe measurements, we subtract the contribution of the resistance from the sheet resistance of the $MoS_2$ and graphene, measured independently (see Supplementary Information). Figure 5a shows $R_c$ measured as a function of $V_{TG}$ for the left (denoted 3-4) and right (5-6) 1DG contacts to $MoS_2$. They are similar, illustrating the homogeneity of the junctions. The $R_c$ values at high carrier density are consistent with the ones extracted using TLM measurements in larger devices, confirming that $R_c$ is independent of contact width, over at least one order of magnitude scaling in the width. When the $MoS_2$ carrier density decreases, the 4-probe measurements show an increase in $R_c$, as is usual for contacts to 2D semiconductors[13].

We have explored the properties of the 1DG contacts at lower carrier densities using additional 4-probe $MoS_2$-based devices controlled using the Si back gate (with no top gate electrode) and similar dimensions as the one shown in Figure 5a, bottom inset. At low carrier densities (n < 3 × $10^{12}$ cm$^{-2}$), where the high resistance of the $MoS_2$ channel itself (above 1 MΩ per square; Figure 5b, upper inset) dominates the total resistance, the I-V characteristics show a linear behavior at room temperature. Only at lower temperatures below 150 K, is there any nonlinearity indicative of an interface barrier (Figure 5b, lower inset). From Arrhenius plots[39,40] we extract the barrier height ($\Phi_B$) as a function of the back gate voltage ($V_{BG}$) as shown in Figure 5b, main panel. We observe $\Phi_B \sim 4$ meV at $V_{BG} = 60$ V and $\Phi_B \sim 24$ meV at $V_{BG} = 10$ V.

The small value of $\Phi_B$ in our devices is further confirmed by the existence of linear I-V characteristics at different values of $V_{BG}$ at room temperature (Supplementary Information) and additional measurements of the differential conductance at liquid helium temperatures. In Figure 6a we plot the differential conductance for the same device in Figure 5b as a function of $V_{BG}$ and the source-drain voltage across the junction measured in a 4-probe geometry ($V_j$). These measurements were performed at a temperature of $T = 4.2$ K, equivalent to a thermal energy of $3k_BT \sim 1$ meV, where $k_B$ is the Boltzmann constant. The fast decrease of the barrier height with increasing $V_{BG}$ is shown by the rapid shrinking of the low differential conductance region[41,42], as represented by the white region (dI/dV = $10^{-8}$ S) in Figure 6a. The sizeable zero-bias differential conductance for $V_{BG} > 40$ V (Figure 6b) is consistent with a barrier height in the same order as the thermal energy (approximately 1 meV). Both the decrease of $\Phi_B$ with $V_{BG}$ and its value around 1 meV at high $V_{BG}$ obtained by our differential conductance measurements are consistent with the thermal-activation (Arrhenius) measurements shown in Figure 5b.

Our $\Phi_B$ values are smaller than the values obtained for overlapped graphene junctions at similar carrier densities, $\Phi_B \sim 20 - 100$ meV in refs. 24 and 28. The small $\Phi_B$ values in our 1DG contacts are consistent with the low-resistance ohmic behavior discussed in Figure 4, and can be explained by the lack of a van der Waals gap in our edge contact geometry.

**Conclusions**

In conclusion, we report the fabrication of one-dimensional ohmic edge contacts between monolayer graphene and monolayer semiconducting TMDs (specifically $MoS_2$ and $WS_2$) using a scalable method. These contacts possess low resistance while maintaining minimal electrode volume and contact area. Our technique for making edge contacts to semiconducting TMDs provides a versatile, stable, and scalable method for forming low-volume, low-resistance contacts for atomically-thin circuitry, which could be attractive for flexible and optically transparent electronics.

**METHODS**

**Heterostructure growth**

Graphene grown by chemical vapor deposition (CVD) on copper is wet-transferred to a $SiO_2$/Si substrate and then patterned using photolithography and oxygen plasma etching. The TMD growth is done using the metal-organic CVD method with Molybdenum hexacarbonyl ($Mo(CO)_6$, MHC, Sigma Aldrich 577766), tungsten hexacarbonyl ($W(CO)_6$, THC, Sigma Aldrich 472956) and diethyl sulphide ($C_4H_{10}S$, DES, Sigma Aldrich 107247) as the chemical precursors for Mo, W, and S respectively. The growth was performed under a temperature of 500 °C and growth time of 30 hours. The precursor vapor pressures are controlled by careful heating of the precursor source and the flow is controlled by mass-flow controllers with settings: 0.01 sccm for MHC or THC, 0.3 sccm for DES, 1 sccm for $H_2$, and 150 sccm for Ar. More detailed information on the heterostructure growth including the ideal experimental settings for junction formation can be found in the supplementary information.

**Device preparation**

After the graphene/TMD lateral heterostructures are grown, a series of lithography steps followed by high-vacuum metal deposition are used to define Ti/Au (5/50 nm thick) electrodes either on the graphene or directly onto the TMD layer. Finally, on some devices, high quality $HfO_2$ (30 to 60 nm) is deposited by atomic layer deposition followed by another lithography and metal deposition step to define the top gate (TG) electrodes. More details can be found in the Supplementary Information.

**Associated Content**

**Supporting Information**. More information on the heterostructure growth and characterization, device fabrication, temperature dependence measurements, and low-temperature differential conductance measurements. This material is available free of charge via the Internet at http://pubs.acs.org.

**AUTHOR INFORMATION**

**Corresponding Author**

*Jiwoong Park (jpark@cornell.edu).

**Author Contributions**

‡ M.H.D.G. and H.G. contributed equally to this work.

**Notes**

The authors declare no competing financial interest.

## ACKNOWLEDGMENTS

This work was supported mainly by the AFOSR (FA2386-13-1-4118), the Netherlands Organization for Scientific Research (NWO Rubicon 680-50-1311), and the Kavli Institute at Cornell for Nanoscale Science. Additional support was provided by the National Science Foundation (NSF DMR-1406333), the Cornell Center for Materials Research (NSF DMR-1120296) and the Nano Material Technology Development Program through the National Research Foundation of Korea (NRF) funded by the Ministry of Science, ICT, and Future Planning (2012M3A7B4049887). Sample fabrication was performed at the Cornell Nanoscale Science & Technology Facility, a member of the National Nanotechnology Infrastructure Network, which is supported by the National Science Foundation (ECCS-1542081).

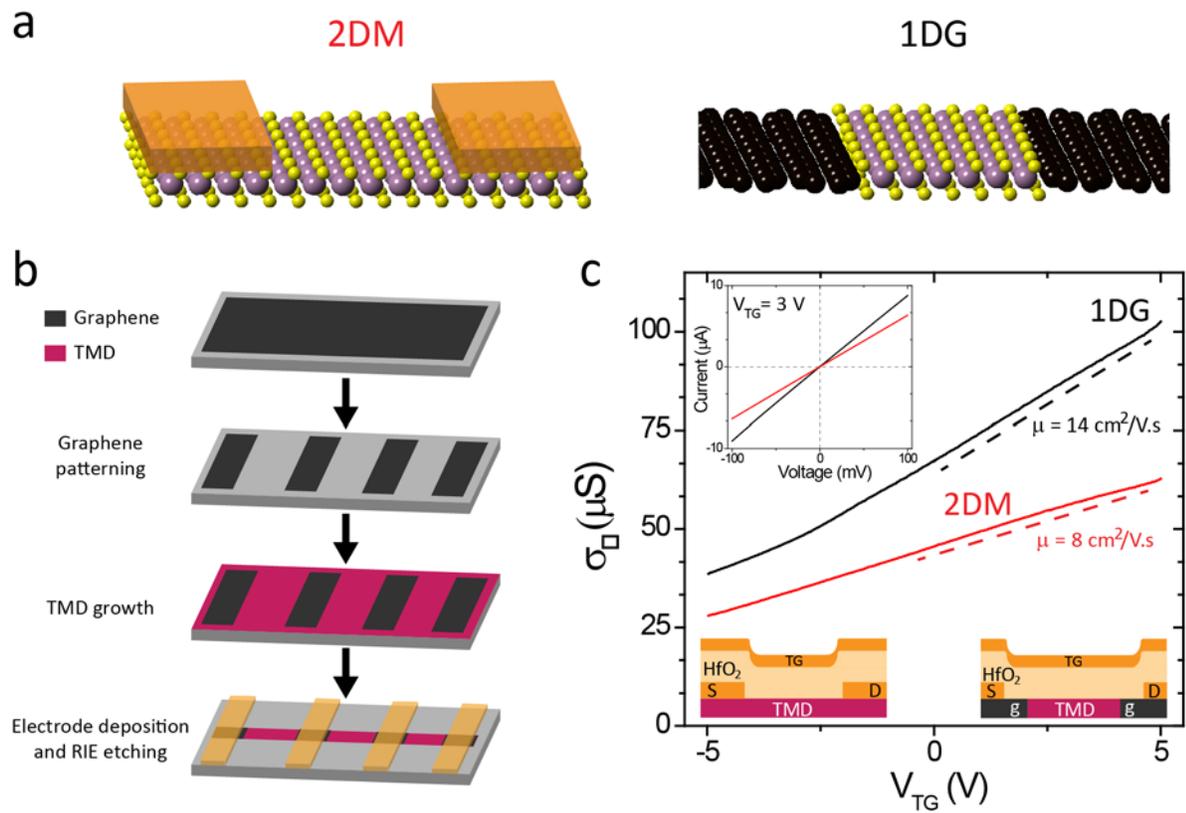

**Figure 1.** (a) Schematic illustration of metal top (2DM) contacts (left) and graphene edge (1DG) contacts (right) for TMD devices. (b) The large-scale growth and fabrication process for producing TMD transistor arrays with 1DG edge contacts. (c) Top gate voltage ($V_{TG}$) dependence of the two-terminal sheet conductance ($\sigma_\square$) measured from MoS$_2$ devices (length 22 μm and width 20 μm) with 1DG contacts (black curve) and with 2DM electrodes (red curve; contact dimensions 23 μm × 22 μm × 55 nm). *Top inset*: IV characteristics for the two devices at $V_{TG}$ = 3 V. *Bottom insets*: Cross-sectional device schematics for the devices with 2DM (left) and 1DG (right) contacts, showing the TMD channel, the graphene contacts (g), and the source (S), drain (D) and top gate (TG) electrodes with the insulating HfO$_2$ layer.

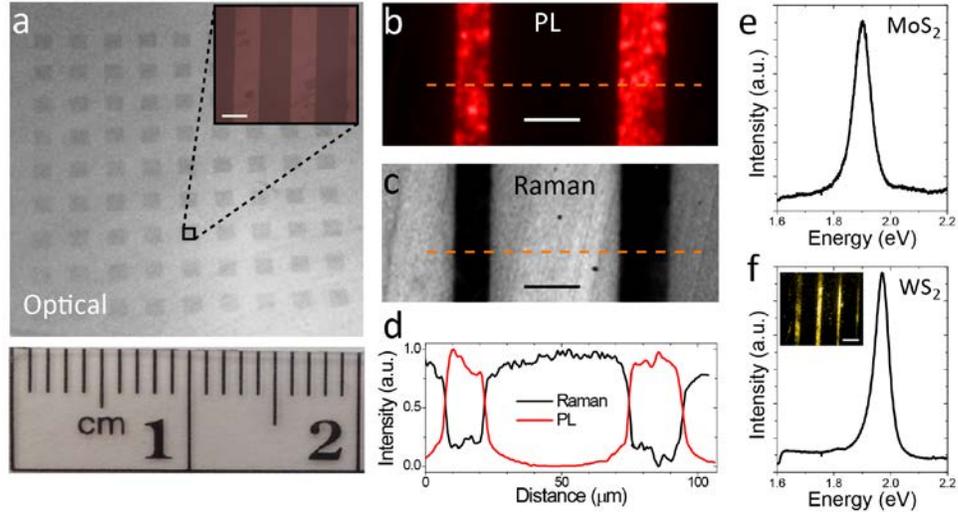

**Figure 2.** (a) Optical image of a typical growth substrate over a 2 cm × 2 cm area. Each dark grey square contains 20 graphene stripes connected by a monolayer $MoS_2$ film. *Inset:* Optical image showing monolayer $MoS_2$ (darker) grown between graphene stripes (lighter). The scale bar is 15 μm. (b) $MoS_2$ photoluminescence (PL) intensity mapping centered at 650 nm and (c) graphene G-band Raman mapping. The scale bars are 25 μm. (d) PL and Raman intensity profiles extracted along the dashed line as indicated in (b) and (c). PL spectrum for (e) $MoS_2$ and (f) $WS_2$. *Inset*: $WS_2$ PL intensity mapping. The scale bar is 50 μm.

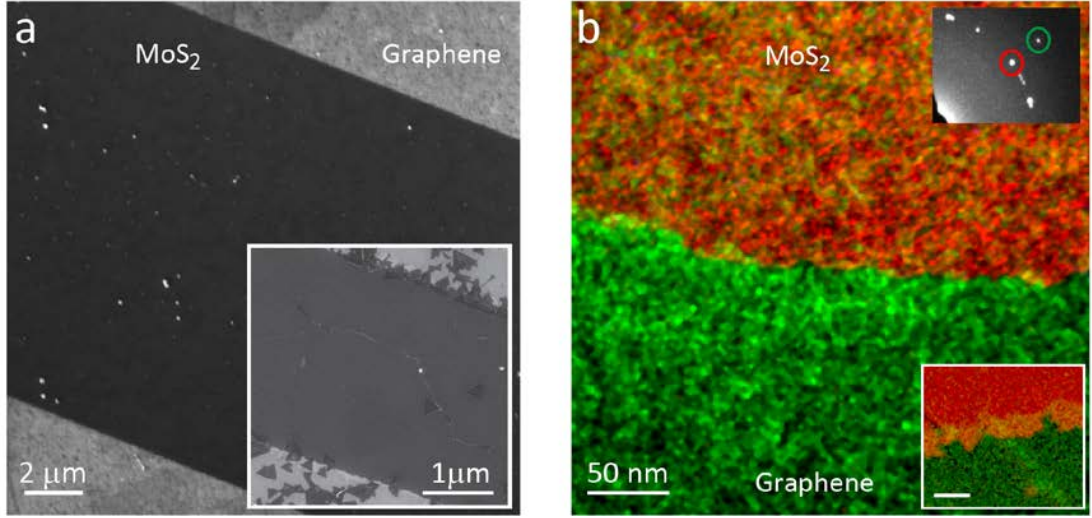

**Figure 3.** (a) Scanning Electron Microscopy (SEM) image of a graphene/MoS$_2$/graphene lateral junction grown under a low Mo precursor pressure. *Inset:* SEM image of a junction grown under high Mo precursor pressure. The presence of few-layer MoS$_2$ flakes (dark triangles) on the graphene and on the monolayer MoS$_2$ film indicate a more reactive growth environment. (b) A dark-field transmission electron microscopy (DF-TEM) image of the lateral junction formed between graphene and MoS$_2$, grown under a low Mo precursor pressure. The *upper inset* shows the diffraction spots used for the graphene area (green) and MoS$_2$ (red). *Lower inset*: A representative DF-TEM image of an overlapped junction obtained under non-optimal (high precursor pressure) growth conditions. The scale bar is 50 nm.

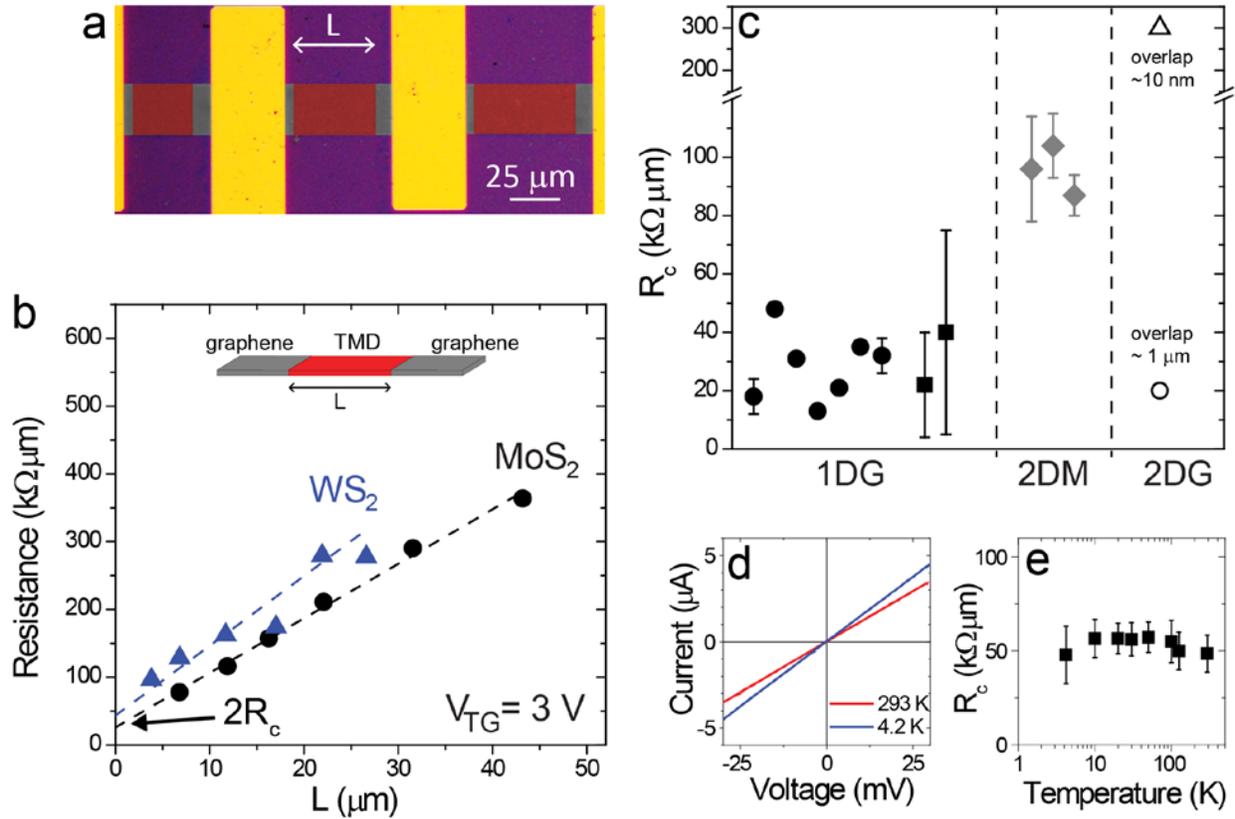

**Figure 4.** (a) Optical micrograph of the device geometry used in the transfer length measurements, where $L$ is the $MoS_2$ channel length. (b) 2-probe resistance as a function of $L$ for $MoS_2$ (black circles) and $WS_2$ (blue triangles) with 1DG contacts. The $y$-intercept of the linear fits gives $2R_c$ and the slope gives the sheet resistance. (c) $R_c$ values for different devices at high carrier density. The solid circles (solid squares) represent our $MoS_2$ ($WS_2$) based devices with 1DG contacts. The gray diamonds represent $R_c$ for our 2DM contacts $MoS_2$. Devices with 2DG contacts from references [28] and [24] are denoted by the open triangle and open circle, respectively. (d) Source-drain current versus voltage at 293 K (red) and 4.2 K (blue) for a $MoS_2$ based device with 1DG contacts at $V_{TG} = 3$ V (n ~ $1 \times 10^{13}$ cm$^{-2}$). (e) $R_c$ as a function of temperature for the same device in (d).

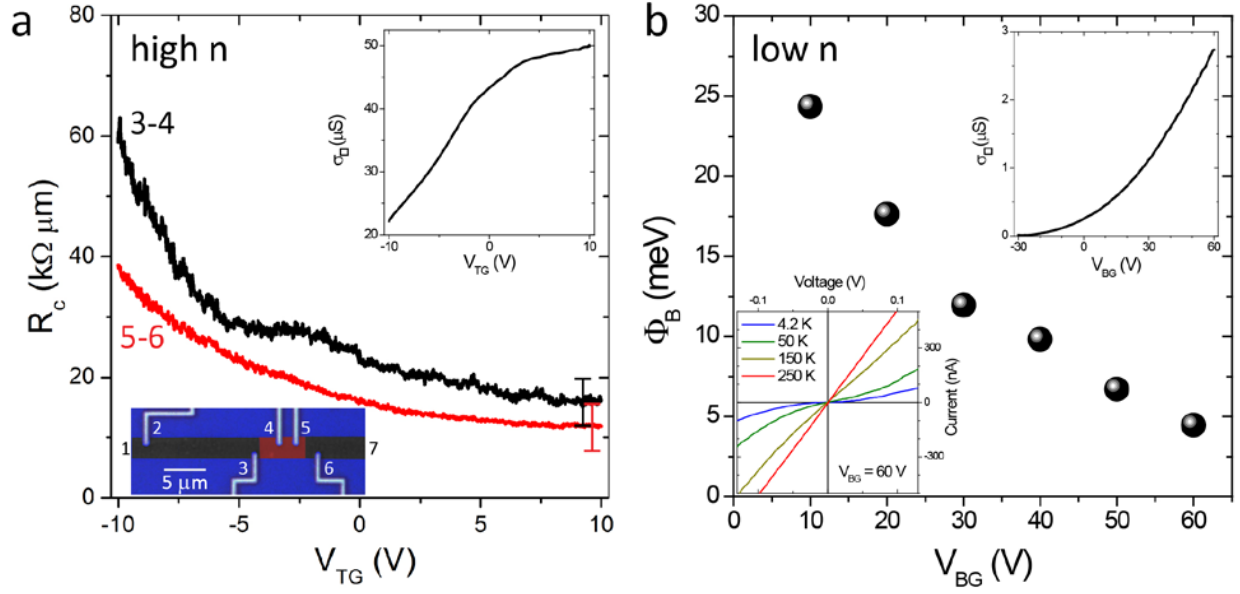

**Figure 5.** (a) Contact resistance $R_c$ as a function of the top gate voltage ($V_{TG}$) for two adjacent junctions at high carrier densities (n > 1 × $10^{13}$ cm$^{-2}$). A false color optical micrograph indicating the electrode numbering is shown in the bottom inset. Current is applied between contacts 1 and 7 and the contact resistances of two different graphene/MoS$_2$ interfaces are measured by reading out the voltage between electrodes 3 and 4 (black points) and 5 and 6 (red points). The contributions from the graphene (between electrodes 2 and 3) and MoS$_2$ (between electrodes 4 and 5) resistances have been subtracted as discussed in the text. *Top inset:* MoS$_2$ sheet conductance measured between electrodes 4 and 5 as a function of $V_{TG}$. (b) Barrier height ($\Phi_B$) as a function of back gate voltage ($V_{BG}$) extracted from the temperature dependence of the junction resistance for a device at low carrier densities (n < 3 × $10^{12}$ cm$^{-2}$). *Top inset:* MoS$_2$ sheet conductance as a function of $V_{BG}$. *Bottom inset:* IV characteristics measured at different temperatures for a single graphene/MoS$_2$ junction in a 4-probe geometry for $V_{BG}$ = 60 V showing nonlinear characteristics due to tunneling below 150 K.

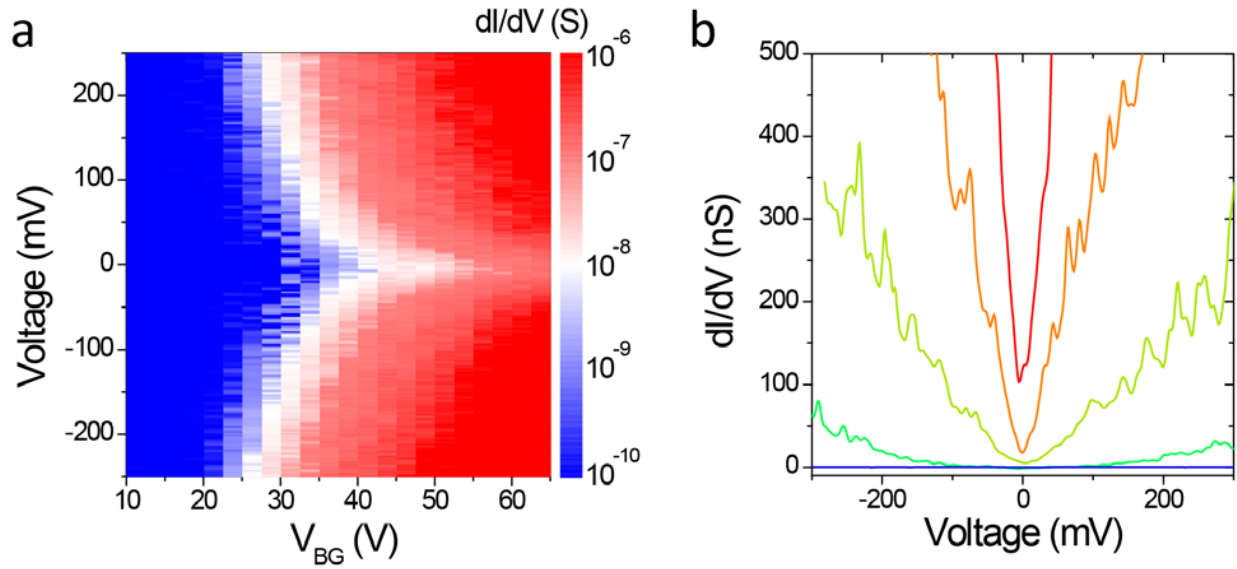

**Figure 6.** (a) Differential conductance ($dI/dV$; log scale) as a function of the back gate voltage ($V_{BG}$) and the source-drain voltage measured across the junction in a 4-probe geometry at low temperature (4.2 K) for the same device as in Figure 5b. (b) $dI/dV$ as a function of the junction voltage for different values of $V_{BG}$ = 20, 30, 40, 50, and 60 V (from bottom to top).